\documentstyle[12pt]{article}

\begin{document}
\addtolength{\baselineskip}{2.3mm}

%----------------------------------------

\begin{center}
{\Large Reply to Comment: \\
Quantum Cryptography Based on Orthogonal States?}
\end{center}
\bigskip

%----------------------------------------

Peres \cite{Peres} claims that our protocol \cite{GV} does not present any 
novel feature, and it is very similar to the oldest protocol of Bennett and 
Brassard \cite{BB84} (BB84). We completely disagree with this claim and with 
other points raised in the Comment. 
The essential novelty of our protocol is that the carrier of information is in 
a quantum state belonging to a definite set of orthonormal states. Any other 
protocol, as well as the BB84 scheme, does not have this feature, and in fact 
their security is based on that. Let us quote a recent paper \cite{EHPP}
stating that:
\begin{quotation}
``...  cloning can give a faithful replica, while leaving the state of the 
original intact, only if it is known in advance that the carrier of 
information is in a quantum state belonging to a definite set of orthonormal 
states. If this is not the case, the eavesdropper will not be able to 
construct even an imperfect cloning device, which would give some information 
on the carrier without modifying it: a device of this sort would violate 
unitarity. Therefore coding based on nonorthogonal quantum states (which 
cannot be cloned) gives the possibility to detect any eavesdropping attempt''.
\end{quotation}
Thus, the security of BB84 (which uses 4 states, not all orthogonal) is 
assured by the `no-cloning' theorem, which is not applicable to our case. 

Peres claims that in our method Eve has access only to nonorthogonal states: 
``The $\rho'_\pm$ states, {\it as seen by Eve\/}, are not orthogonal. Their 
are {\it identical\/}". However, the nonorthogonality (as seen by Eve) in our 
scheme is not ``just as in the BB84 protocol''. As Peres admits, in the case 
of {\it known} sending times our protocol is not secure, yet his 
nonorthogonality argument remains the same. The security of our protocol is 
not based on nonorthogonality, but on causality. As we have proved in the 
Letter \cite{GV}, a successful eavesdropping is possible only if some 
information can reach Bob before it leaves Alice's site - therefore, the 
protocol is secure. This is also the feature of the protocol proposed at the 
end of the Comment, in which a particle is sent at a known time in one out of 
two GV interferometers (GV2). A successful eavesdropping is impossible in this 
case too, otherwise the wavepacket delayed at Alice's location has to reach 
Bob's site before leaving Alice' site. Note that in the case of a single GV 
interferometer (and known sending times) Eve can send Bob a dummy particle at 
the appropriate time, but here she does not know which of the two 
interferometers to use for sending it.

According to Peres, an important common feature of GV and BB84 (or other 
protocols interpolating between these two) is that ``information is sent in 
two {\it consecutive} steps, and security is achieved by withholding the 
second piece of information until after Bob receives the first one''. In his 
view the first step is sending the particle, and the second step is sending 
the necessary classical information: the chosen basis (BB84), the transmission 
time (GV), or the chosen interferometer (GV2). The first conceptual difference 
between the protocols is that in BB84 the two steps are necessary for sending 
the information, while in GV or GV2 one step is enough. The only purpose of 
the second step is to assure security against eavesdropping. 
The second difference is that the first step of our protocol also consists of 
two stages: sending the first wavepacket and sending the second wavepacket 
(the delayed one). Alice does not have to wait until the end of the first step 
for announcing the sending time (GV) or the interferometer in use (GV2). She 
can do that after the first stage of the first step (i.e. after the first 
wavepacket reaches Bob), thus, `the second step' might end before `the first 
step'. These two stages of the first step, i.e. the fact that the quantum 
signal consists of two separated parts, is the core of our method, and we do 
not see its analog in BB84 or any other protocol.

Finally, it seems that Peres has not understood the `relativistic' versions 
of our protocol. First, it is not true that the storage rings have to be 
larger than the distance between Alice and Bob. When the communication is 
based on photons which travel on straight lines, the time delay can be made as 
small as wanted (it depends on the width of the wavepackets and on the 
accuracy of the clocks). Contrary to Peres' claim (and his Fig. 1), Eve can 
simultaneously access the two branches of the interferometer most of the time, 
still the protocol is secure. A similar proof to that given in the Letter 
\cite{GV} shows that a successful eavesdropping leads to superluminal 
signaling.
Second, it is not true that in the case of the protocol with two widely 
separated paths and no time delay, ``a {\it team} of eavesdroppers could use 
mirrors to redirect the photon paths toward a common inspection center, and 
hence to Bob, without arousing suspicion''. Such an operation invariably 
increases the flight-time of the photons, therefore the users can easily 
expose the team by analyzing the timing. Since the information is encoded in 
the relative phase between the wavepackets, even more sophisticated 
eavesdropping methods cannot work, unless they use superluminal particles.

%----------------------------------------

\bigskip
\noindent
Lior Goldenberg and Lev Vaidman \\
School of Physics and Astronomy, \\
Raymond and Beverly Sackler Faculty of Exact Sciences, \\
Tel-Aviv University, \\
Tel-Aviv 69978, Israel 

%----------------------------------------

%----------------------------------------

\end{document}